\documentclass{ws-procs9x6}

\newcommand{\lo}{leading order~}

\newcommand{\nlo}{next-to-leading order~}
\newcommand{\nloo}{next-to-leading order}

\def\Journal#1#2#3#4{{#1} {\bf #2}, #3 (#4)}  
  
\def\NPB{{\em Nucl. Phys.} B}  
\def\PLB{{\em Phys. Lett.}  B}  
  
\def\PRD{{\em Phys. Rev.} D}

\begin{document}


\title{Spin-Flavor Decomposition 
in Polarized Semi-Inclusive Deep Inelastic Scattering Experiments at Jefferson Lab}

\author{Xiaodong Jiang \footnote{\uppercase{W}ork 
supported by the \uppercase{U.S.}
\uppercase{N}ational \uppercase{S}cience \uppercase{F}oundation grant \uppercase{NSF}-\uppercase{PHY}-03-54871.}}

\address{Department of Physics and Astronomy, \\
Rutgers, the State University of New Jersey, \\
136 Frelinghuysen Road,  Piscataway, NJ 08854 USA.\\ 
E-mail: jiang@jlab.org}

\maketitle

\abstracts{ A Jefferson Lab experiment proposal was discussed in this talk. 
The experiment is designed to measure the beam-target double-spin 
asymmetries $A_{1n}^h$ in semi-inclusive deep-inelastic $\vec n({\vec e}, e^\prime \pi^+)X$ 
and $\vec n({\vec e}, e^\prime \pi^-)X$ reactions 
on a longitudinally polarized $^3$He target.
In addition to $A_{1n}^h$, the flavor non-singlet combination $A_{1n}^{\pi^+ - \pi^-}$, 
in which the gluons do not contribute, will be determined with high precision to extract 
$\Delta d_v(x)$ independent of the knowledge of the fragmentation functions.
The data will also impose strong constraints on quark and gluon polarizations through a global NLO
QCD fit. 
}

\section{Introduction}
Polarized semi-inclusive deep-inelastic scattering (SIDIS) experiments 
can be used 
to study the spin-flavor structure 
of the nucleon, as has been demonstrated first by the SMC experiment~\cite{smc1998}.
Recently, the HERMES experiment
published results of a leading order spin-flavor decomposition from polarized 
proton and deuteron SIDIS asymmetry data, and for the first time
extracted the $\bar{u}, \bar{d}$ and $s=\bar{s}$ sea quark polarizations~\cite{hermes2004}. 
The HERMES  ``purity'' method of spin-flavor decomposition 
relies on the assumption that
the quark fragmentation process and the experimental 
phase spaces are well-understood such that a LUND model 
based Monte Carlo simulation can reliably reproduce the probability correlations
between the detected hadrons and the struck quarks~\cite{hermes2004}. The accuracies on the
knowledge of the fragmentation process played a crucial role in extracting the 
polarized parton distributions in this method.  

As an alternate method, Christova and Leader pointed out~\cite{leader2} that 
if the flavor non-singlet combination $A_{1}^{\pi^+ - \pi^-}$ is measured,
the quark polarization $\Delta u_v$, $\Delta d_v$ and $\Delta \bar{u} - \Delta \bar{d}$ can be extracted
at  \lo without the complication of fragmentation functions.  In fact,
 information on the valence quark polarizations will be well preserved at any QCD order 
in $A_{1}^{\pi^+ - \pi^-}$ since gluons do not contribute to this 
flavor non-singlet observable. At leading order, 
assuming isospin symmetry and charge conjugation, the fragmentation functions 
cancel exactly in $A_{1}^{\pi^+ - \pi^-}$ and
the $s$-quarks do not contribute, so that:
\begin{eqnarray}
\label{Eq:cl1}
&A&\hspace{-0.1cm}_{1p}^{\pi^+ - \pi^-}  \equiv  { \Delta \sigma_p^{\pi^+}-\Delta \sigma_p^{\pi^-} \over
\sigma_p^{\pi^+} - \sigma_p^{\pi^-} }= {A_{1p}^{\pi^+} - A_{1p}^{\pi^-} \cdot \sigma_p^{\pi^-}/\sigma_p^{\pi^+} \over
1-\sigma_p^{\pi^-}/\sigma_p^{\pi^+}}=
{  4\Delta u_v - \Delta d_v 
\over 4u_v - d_v },  \nonumber \\
&A&\hspace{-0.1cm}_{1n}^{\pi^+ - \pi^-}   \equiv { \Delta \sigma_n^{\pi^+}-\Delta \sigma_n^{\pi^-} \over
\sigma_n^{\pi^+}- \sigma_n^{\pi^-} }= {A_{1n}^{\pi^+} - A_{1n}^{\pi^-} \cdot \sigma_n^{\pi^-}/\sigma_n^{\pi^+} \over
1-\sigma_n^{\pi^-}/\sigma_n^{\pi^+}}=
{ 4 \Delta d_v - \Delta u_v 
\over 4 d_v - u_v}.
\end{eqnarray}
Thus, measurements of $A_{1}^{\pi^+ - \pi^-}$ on the proton and 
the neutron can determine $\Delta u_v$ and $\Delta d_v$. 
On the other hand, another non-singlet quantity is constrained by the inclusive data:
\begin{equation}
\label{Eq:nlog1pn}
g_1^p(x,Q^2) - g_1^n(x,Q^2) = { 1 \over 6 } \left[ (\Delta u + \Delta \bar{u}) - (\Delta d + \Delta \bar{d}) \right] 
\vert_{LO}.
\end{equation}
If one is only interested in flavor non-singlet quantities, such as $\Delta \bar{u} - \Delta \bar{d}$,
the goal of SIDIS experiments is reduced to obtaining information on $\Delta u_v -\Delta d_v$, and
the polarized sea asymmetry can be extracted at \lo:
\begin{equation}
\label{Eq:dubardbar}
(\Delta \bar{u} - \Delta \bar{d}) \vert_{LO} = 3 (g_1^p- g_1^n)\vert_{LO} 
 - {1 \over 2} (\Delta u_v - \Delta d_v) \vert_{LO}.
\end{equation}

At  \nloo, QCD global fits of
data from both inclusive and semi-inclusive reactions
have become the state of the art~\cite{sassotnlo}.   One expects that the next generation 
 of  polarized parton distribution functions will take advantage 
of improvements on both quality and volume of SIDIS data from HERMES, COMPASS and Jefferson Lab.
 Although there are data available 
and experiments planned~\cite{e04113} with polarized proton and deuteron targets,  
there has been a lack of attention in obtaining SIDIS data on the 
neutron from a polarized $^3$He target.    

\section{A proposal of polarized $^3$He SIDIS at Jefferson Lab}

An experiment proposal~\cite{p05112} has been developed recently at Jefferson Lab 
Hall A to provide high statistics neutron
SIDIS data using a polarized $^3$He target. 
The plan is to measure the double-spin asymmetries $A_{1n}^h$ in ${\vec n}({\vec e},e^{\prime}h)X$ 
reactions ($h= \pi^+$ and $\pi^-$, with $K^+$ and $K^-$ as by-products)  
with a 6 GeV polarized electron beam on 
a longitudinally polarized $^3$He target at a luminosity of 10$^{36}$ sec$^{-1}$cm$^{-2}$.
The Hall A left-HRS spectrometer with its septum magnet will 
 detect the leading hadrons at $6^\circ$ ($\Delta \Omega \approx 5$ msr)
with a momentum of 2.40 GeV/c ($z_h=E_h/\nu \sim 0.5$)
for either positive or negative polarity.
The recently constructed BigBite spectrometer ($\Delta \Omega \approx 60$ msr) will be used as
the electron detector at 30$^\circ$ to detect the scattered electrons with
 $0.8 \sim 2.1$ GeV/c in coincidence
($0.12<x<0.41$, $Q^2=1.21\sim 3.14$ GeV$^2$).
Since the $\pi^-$ and $\pi^+$ phase spaces are identical 
and the detection efficiencies can be well-controlled, relative 
$\pi^-/\pi^+$ yield ratios can be easily determined such that the flavor non-singlet combination
 $A_{1n}^{\pi^{+} - \pi^{-}}$ can be constructed.

\begin{figure}[ht]
\centerline{\epsfxsize=2.75in\epsfbox{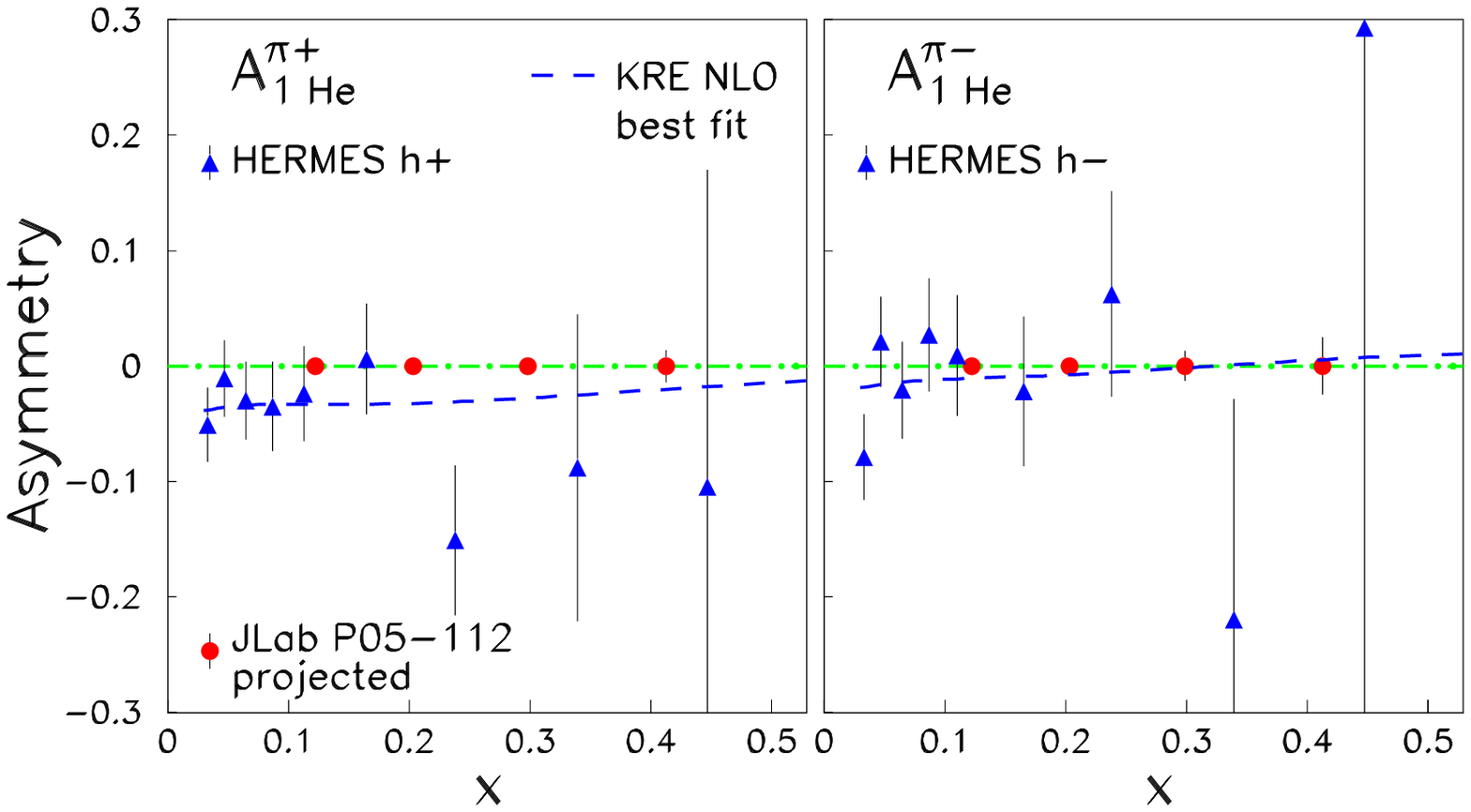}\epsfxsize=2.75in\epsfbox{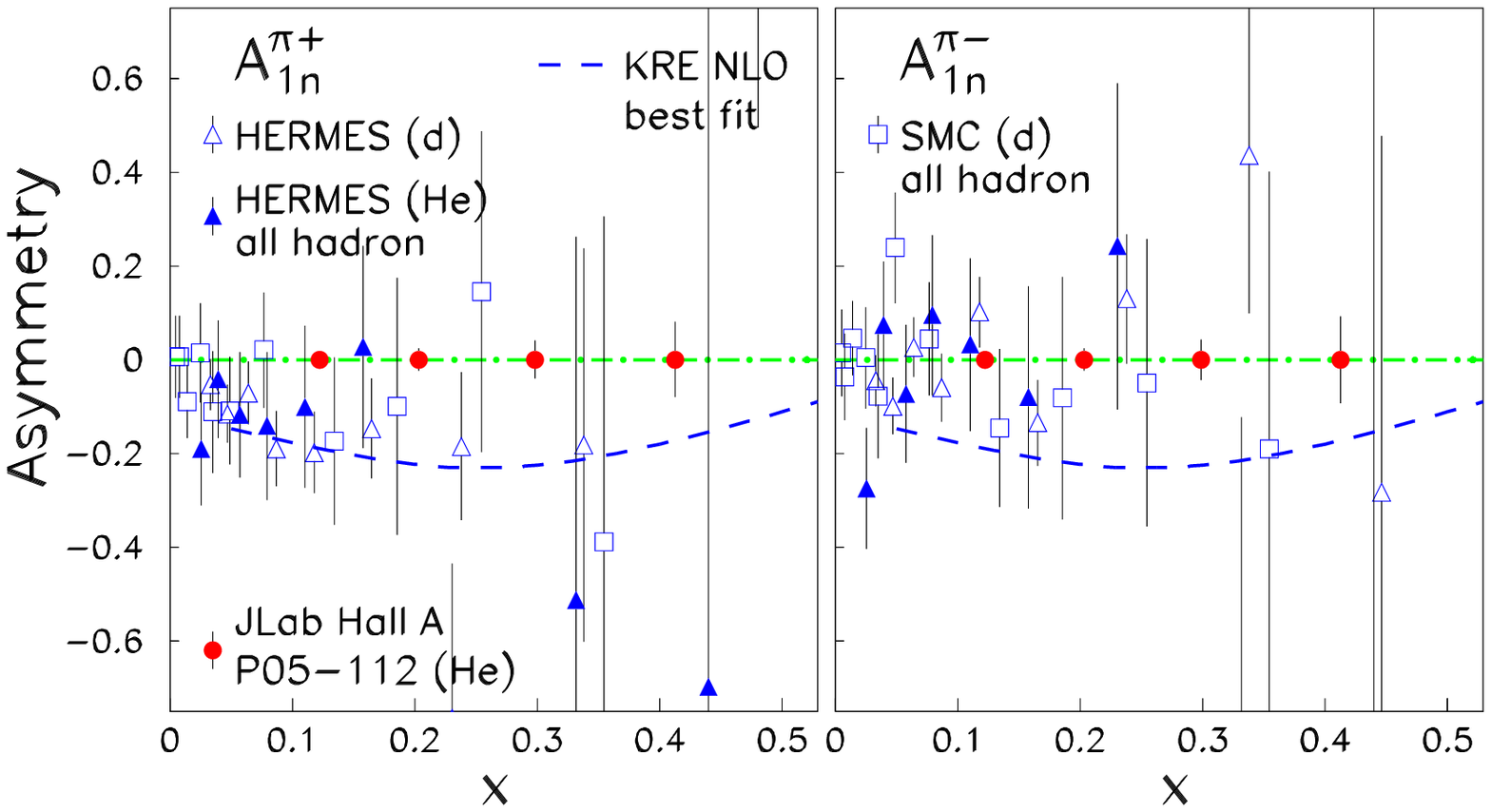}}   
\caption{
 The expected statistical uncertainties of $A_{1He}^\pi$ (left) and $A_{1n}^\pi$ (right) of Jefferson Lab proposal P05-112. 
 The SMC and the HERMES deuteron data~\protect\cite{smc1998,hermes2004}  
 have been translated into neutron asymmetries assuming leading order $x$-$z$ factorization.
 The dashed curves are from the \nlo global fit~\protect\cite{sassotnlo} of the existing data.
\label{fig:a1he3} 
}
\end{figure}

\begin{figure}[ht]
\centerline{\epsfxsize=3.05in\epsfbox{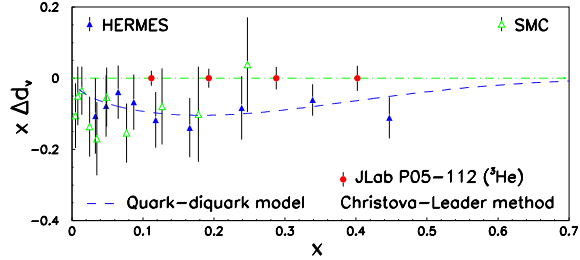}}   
\caption{
The  statistical accuracy of $\Delta d_v$
compared to the SMC~\protect\cite{smc1998} and the HERMES data~\protect\cite{hermes2004}.  
 The dashed curve is from a covariant quark-diquark model calculation~\protect\cite{cloet}
 of Cloet {\it et al.} \label{fig:deltadv}
}
\end{figure}

For 30 days of beam time, assuming 75\% beam polarization and 42\% target polarization, the statistical 
accuracy of $A_{1He}^\pi$ can be improved by an order of magnitude 
compared to earlier HERMES data~\cite{hermes1999}, as shown in Fig.~\ref{fig:a1he3}. Significant improvements are expected on $A_{1n}^\pi$ 
when compared to the deuteron data from SMC and HERMES.
Following the Christova-Leader method of 
Eq.~\ref{Eq:cl1}, $\Delta d_v$ can be extracted from $A_{1n}^{\pi^+ -\pi^-}$,
 as shown together in Fig.~\ref{fig:deltadv}
 with the HERMES data~\cite{hermes2004} from the purity method. 
When combined with the upcoming proton data~\cite{e04113} of JLab experiment E04-113, 
sea flavor asymmetries can be extracted at \lo
following Eq.~\ref{Eq:dubardbar}. With a factor of five improvement on statistical accuracy  
compared to that of the HERMES $\Delta \bar{u}- \Delta \bar{d}$ results,  this 
experiment might provide the first opportunity to discover a possible polarized sea asymmetry.

\begin{figure}[htb]
\centerline{\epsfxsize=3.75in\epsfbox{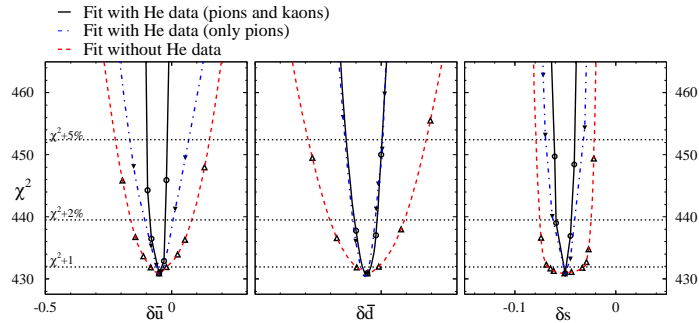}}   
\caption{The expected improvement on the sea polarization moments in NLO global fit~\protect\cite{sassotnlo}. 
The dashed lines are the existing constraints, the dot-dashed and the solid lines are 
the constraints after adding pion data and/or kaon data from this proposal.
The horizontal lines correspond to a deviation of $\chi^2+1$, 
$\chi^2(1+2\%)$ and $\chi^2(1+5\%)$ from the best fit. \label{fig:deltaqb}
}
\end{figure}

\begin{figure}[htb]
\centerline{\epsfxsize=2.05in\epsfbox{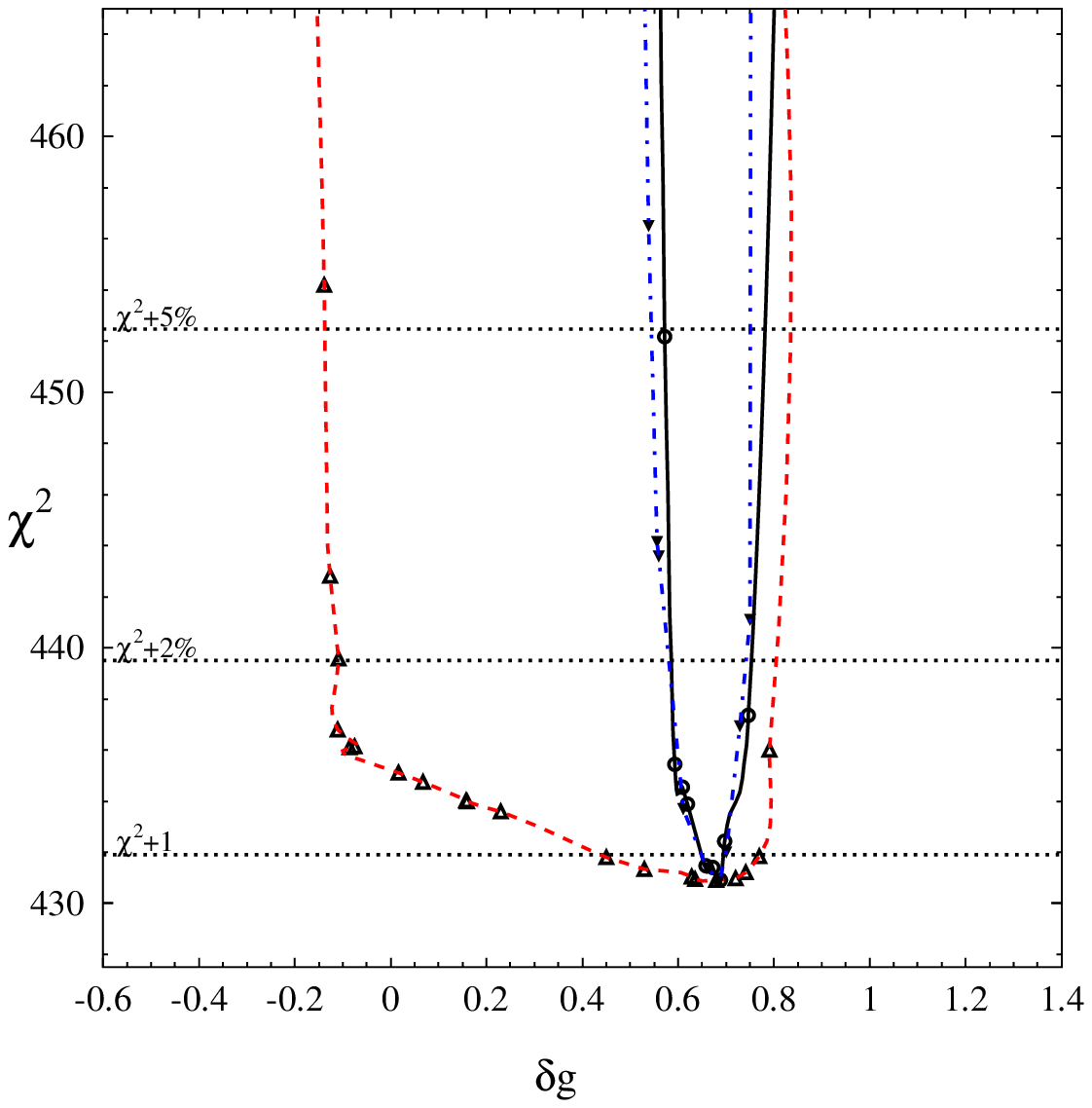} \epsfxsize=2.30in\epsfbox{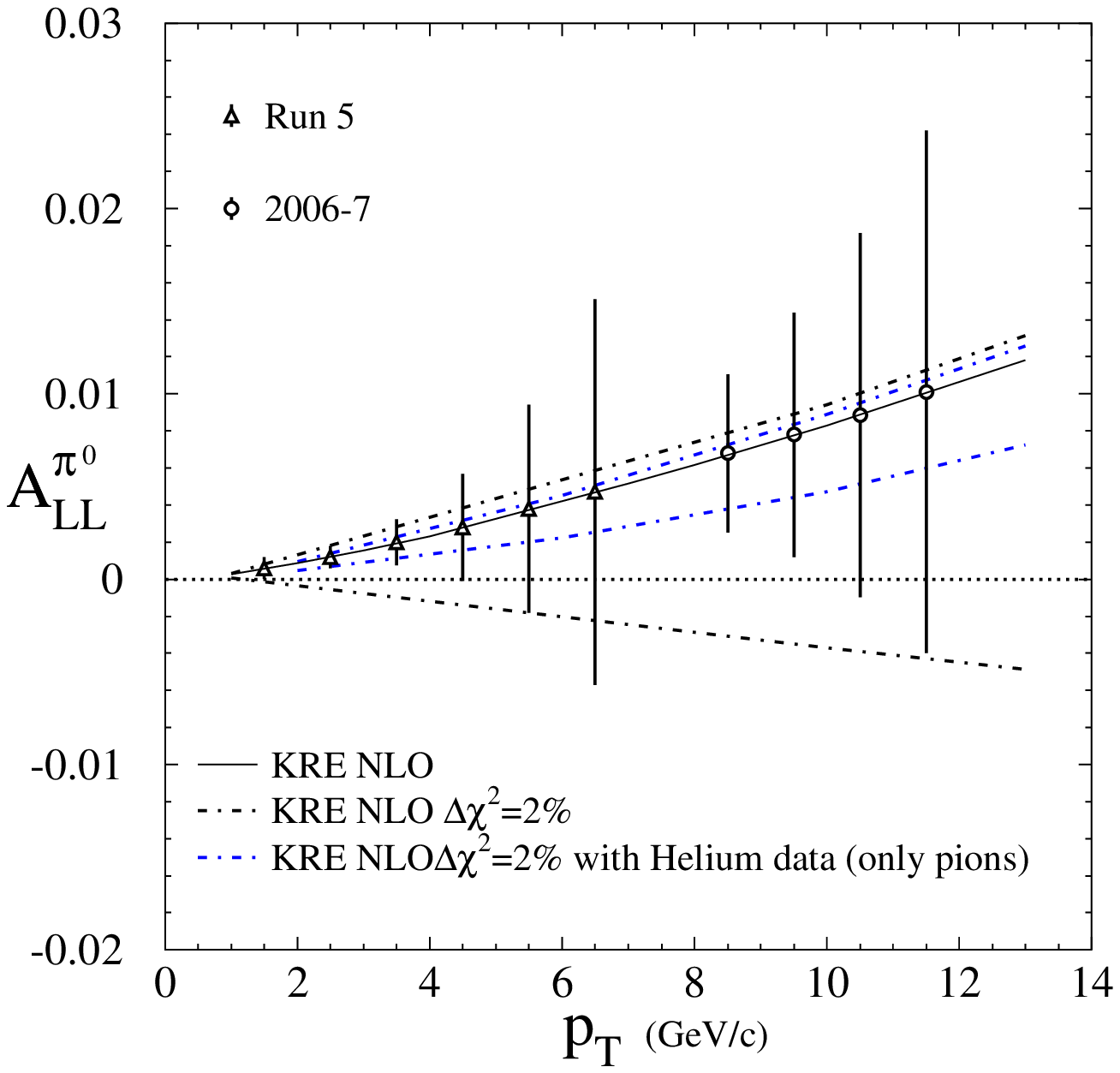}}   
\caption{The constraint on the moment of the gluon polarization~\protect\cite{sassotnlo}
by this measurement (left, curves as labled in Fig.~\protect\ref{fig:deltaqb}) is compared with that from the expected RHIC-2007 $A_{LL}^{\pi^0}$ data (right). 
In the right panel, the area covered  between the 
two inner blue dot-dashed lines corresponds to the $\chi^2(1+ 2\%)$ crossover region
 with the blue-dashed line on the left panel.
 \label{fig:deltag}
}
\end{figure}

Adding this set of $^3$He data to the global NLO QCD fit, a factor of three improvement on 
sea quark polarization moments can be expected~\cite{sassotnlo}, as shown in 
Fig.~\ref{fig:deltaqb}. Indirectly, this data set will also improve the constraints on the 
gluon polarization 
$\Delta g$ by a factor of three, comparable to the impact of the expected RHIC-2007 $A_{LL}^{\pi^0}$ data, as shown 
in Fig.~\ref{fig:deltag}. 
The reason for this sensitivity is because $\Delta g$ 
is obtained in the global fit through 
the $Q^2$-evolutions of the inclusive structure functions $g_1$ which are coupled to the sea distribution. 
Once the valence distribution is reasonably separated from the sea with the SIDIS data, 
 the gluon polarization can be better constrained in a global fit.

\vspace{-0.10in}
\section{Conclusions}
A Jefferson Lab Hall A experiment proposal to measure polarized $^3$He 
SIDIS asymmetries in $\vec n({\vec e}, e^\prime h)X$ reactions was discussed. 
The proposed measurement will dramatically improve the precision of 
the world data set of $A_{1n}^h$ and our knowledge of $\Delta d_v(x)$.
Strong constraints on quark and gluon polarizations can be imposed through a global NLO QCD 
analysis.

\vspace{-0.10in}
\section*{Acknowledgments}
The author thanks G.~A.~Navarro, R.~Sassot, E.~Christova, E.~Leader, C.~Weiss, J.-P. Chen, 
R. Gilman and J.-C. Peng
for many discussions.


\vspace{-0.10in}


\begin{thebibliography}{0}
\bibitem{smc1998} The Spin Muon Collaboration, \Journal{\PLB}{420}{180}{1998}.
\bibitem{hermes2004} The HERMES collaboration, 
                     \Journal{\PRD}{71}{012003}{2005}.
\bibitem{leader2} E. Christova and E. Leader, 
                                             \Journal{\NPB}{607}{369}{2001}.
\bibitem{sassotnlo} D.~de Florian, G.~A.~Navarro and R.~Sassot, \Journal{\PRD}{71}{094018}{2005}. 
\bibitem{hermes1999} The HERMES collaboration, \Journal{\PLB}{464}{123}{1999}.
\bibitem{e04113} Jefferson Lab experiment E04-113, X.~Jiang, P.~Bosted, D.~Day and M.~Jones co-spokespersons, hep-ex/0412010.
\bibitem{p05112} Jefferson Lab experiment proposal P05-112, X.~Jiang spokesperson.
\bibitem{cloet} I.C.~Cloet, W.~Bentz and A.W.~Thomas, \Journal{\PLB}{621}{246}{2005}.
\end{thebibliography}
\end{document}